\documentstyle[12pt,epsf]{article}

\begin{document}
\begin{titlepage}

{\hspace*{\fill} SU-4240-606\\
 \hspace*{\fill} PAR/LPTHE/95-27 \\
 \hspace*{\fill} hep-ph/9505294 \\
 \hspace*{\fill} May 1995\\ }

\bigskip\bigskip

\begin{center}
      {\Large\bf Puzzle in the Charmed $D_s$ meson decays into pions: Could
         the light quarks be not so light?\\}
\end{center}

\bigskip

\begin{center}
{\large  Hoang Ngoc Long $^a$ , Nguyen Ai Viet $^{a,b}$
 and Pham Xuan Yem$^c$\\}
\medskip\medskip
{\it $^a$ High Energy Group, Institute of Physics,\\
 P.O.Box.429, Bo Ho, Hanoi 10000, Vietnam\\}
\medskip
{\it $^b$ Physics Department, Syracuse University\\Syracuse
, NY 13244-1130, USA\\}
\medskip
{\it $^c$ Universit\'e  Pierre et Marie Curie, Paris VI \\
   Universit\'e Denis Diderot, Paris VII \\
 Physique Th\'eorique et Hautes \'Energies,\\
( Unit\'e associ\'ee au CNRS : D 0280 ) \\
75252 Paris C\'edex 05, France  }
\end{center}

\bigskip

\begin{abstract}
The inclusive decay rate into pions of the charmed $D_s$ meson is surprisingly
larger than estimates expected from the $W$ annihilation, adopting
commonly used values of current-algebra up and down quark masses. We then go
beyond this tree diagram and consider possible QCD effects that might cause
such a large rate. There are two; the first one is related to the spectator
decay $ c{\bar s}
\rightarrow s{\bar s} + u {\bar d}$ followed by $ s {\bar s} \rightarrow
d{\bar d}~,~ u{\bar u}$ via two-gluon exchange box diagram. The second one is
a gluon emission in weak annihilation for which the usual helicity
suppression is vitiated:
$D_s\rightarrow W+g$ followed by $ W\rightarrow u {\bar d},~ g\rightarrow
d{\bar d},~u{\bar u}$.

These two contributions, however, turn out to be insufficient to explain
data, implying that the puzzle could be understood if the up, down quarks have
higher mass values.

Furthermore, on the basis of experimental informations on the spectral function
$\rho_{3\pi}(Q^2)$ deduced from the exclusive $ D_s \rightarrow 3\pi$ mode,
the QCD sum rules also point to a higher mass for light quarks.
\end{abstract}

\end{titlepage}
The inclusive decay rate of the charmed $ D_s$ meson into non-strange ordinary
hadrons ( mainly pions) is surprisingly large \cite{PDG}. Its amplitude, being
governed by the $ W$ annihilation mechanism at the electroweak tree diagram and
directly related to the
divergence of the axial current, is expected to be strongly suppressed by the
familiar helicity argument ( similar to the suppression of $\pi \rightarrow e
\nu$ compared to $ \pi \rightarrow \mu\nu$), and/or by the partial conservation
of the axial current (PCAC). Experiments do not confirm this expectation,
however.

We observe first that the $ D_s$ decay into pions cannot be described either
by the dominant spectator mechanism ( both color-favoured and
color-suppressed) or by the small penguin diagram, because the spectator
constituent $ {\bar s}$ of the $ D_s^+$ is absent in the decay products.
Therefore only the $ W $ annihilation mechanism can give rise - at the tree
level - to the decays of $ D_s$ into pions. The gluonic effects will be
also considered. Experimentally, the inclusive branching ratio for $ D_s$
decays into
non-strange hadrons ( denoted in the following by $X_{ud}$) can be estimated to
be at least $(1.65 \pm 0.35)\%$ from the Particle Data Group (PDG) \cite{PDG}.
For this lower bound, we only retain the three and five charged pion modes. All
other modes with four, six, seven pions are disregarded because they come
mainly
from the quasi two-body modes $ D_s \rightarrow \eta + \pi(\rho)$ and $ D_s
\rightarrow \eta' +\pi(\rho)$ followed by the subsequent decays of $\eta$ and
$\eta'$ into pions. Since $\eta$ and $\eta'$ have a large $s{\bar s}$
component, these modes must be attributed to the dominant spectator diagram and
not to the $ W $ annihilation in which we are interested here. Let us remark
that  with the $W$ annihilation, the decays of $ D_s$ into $ \eta + \pi$
and $\eta' + \pi$ vanish by the conserved vector current (CVC) in the standard
factorization approach.
Another remarkable feature of the experimental data is
the important fraction of $D_s$ decays
into non-resonant $\pi^+\pi^-\pi^+$ state.
Its rate, which is already around 1/3 of the dominant spectator $\phi \pi^+$
one, is really intriguing.

In the rest of the paper, we use two different methods - appropriated to
inclusive and exclusive decays - to estimate the up, down quark masses. Both
approaches converge to mass values higher than the ones estimated in
literature.

\section{Inclusive Decay of $ D_s$ into pions:}
Compared to the pure leptonic rate $ D_s \rightarrow \mu+\nu$, the inclusive
decay of $D_s$ into non-strange hadrons $ X_{ud}$ is given, at the effective
electroweak tree level, by:
\begin{equation}
R_{D_s} \equiv { \Gamma(D^+_s \rightarrow X_{ud}) \over \Gamma(D^+_s
\rightarrow \mu^+ \nu)}~=~{\Gamma(D^+_s \rightarrow u{\bar d})\over
\Gamma(D^+_s \rightarrow \mu^+\nu)}~=~ 3a^2_1 |V_{ud}|^2 {J(x_u, x_d)\over
(1-x_\mu)^2}{m_u^2+ m_d^2 \over m^2_\mu}.
\end{equation}
By considering the ratio $ R_{D_s}$ in Eq.(1), the large uncertainty in the
decay
constant $ f_{D_s}$ can be avoided. The coefficient $3$ comes from color. Here
$a_1$ is the Bauer-Stech-Wirbel (BSW) phenomenological parameter \cite{BSW}
taken from the QCD corrected effective Lagrangian first calculated by
Gaillard-Lee \cite{GLAM}, Altarelli-Maiani \cite{GLAM} following the Wilson
operator product expansion method \cite{WILSON}
\begin{equation}
{\cal L}_{eff}~=~ {G\over \sqrt{2}} V^*_{cs} V_{ud} [~ c_1( {\bar s}
\gamma_{\mu
L} c)( {\bar u} \gamma_{\mu L} d) + c_2({\bar s}\gamma_{\mu L} d)({\bar
u}\gamma_{\mu L} c)~]
\end{equation}
In the vacuum insertion approximation, known as the factorization method \`a
la BSW \cite{BSW}, the relevant effective Lagrangian appropriated to our case
can be written as \cite{BSW}
\begin{equation}
{\cal L}_{eff} ~=~ {G\over \sqrt{2}} V_{cs}^*V_{ud} a_1 ({\bar s} \gamma_{\mu
L} c)_H({\bar u}\gamma_{\mu L} d)_H
\end{equation}
with $ a_1= c_1 + {1\over N_c} c_2 $ and the subscript $H$ stands for color
singlet hadronic currents \cite{BSW}.

 From the general fit of charm non-leptonic decays, the value of $ a_1 = 1.26$
is commonly used \cite{5}.

Finally $ J(x_u, x_d)$ in Eq.(1) is the phase-space correction factor \cite{6}
due to the quark masses
\begin{equation}
J(x_u, x_d) ~=~ \big [1- {(x_d-x_u)^2 \over x_d+x_u}\big ]~ \lambda(1,x_u,x_d)
\end{equation}
with
$ \lambda (1, x, y) ~=~ \sqrt{ (1-x-y)^2- 4 x y},~~ ~~ x_{u,d,\mu} =
{m^2_{u,d,\mu} \over M^2}$
where
$ m_u, m_d, m_\mu$
and
$ M$
are respectively
masses of up, down quarks, muon and charmed $D_s$ meson.

We remark the similarity of the ratio $R_{D_s}$ with the hadrons/lepton ratio $
R_\tau $ in the tau's lepton decay
\begin{equation}
R_\tau \equiv {\Gamma( \tau^+ \rightarrow {\bar \nu}_\tau X_{ud}) \over
\Gamma( \tau^+ \rightarrow {\bar \nu}_\tau e^+ \nu)} = 3 |V_{ud}|^2 F(x_u, x_d)
\end{equation}
with the phase space factor \cite{6} $ F(x_u, x_d)$ given by
\begin{eqnarray}
F(x, y) &=& \lambda(1, x, y) [ 1- 7(x+y) -(x^2+y^2)-6(x-y)^2+(x+y)(x-y)^2
 \nonumber\\
 &-& 6xy(x+y)]+ 12[ x^2(1-y^2) ~ln{1+x-y+\lambda(1, x, y) \over 1+x-y -
\lambda(1,x, y)} + (x \leftrightarrow y)]
\end{eqnarray}
$ J(0,0) ~=~ F(0,0) ~= 1$.

Eqs.(1) and (5) are only exact at the tree level, i.e. when QCD effects - at
the
final up, down vertex- are not taken into account. When gluonic effects ( up to
three loops) at the final quarks vertex are kept \footnote{ Let us emphasize
that
there is no confusion ( or double counting) possible between the QCD correction
in the effective Lagrangian Eq.(2) symbolized by the $ c_{1,2}$ coefficients on
the one hand, and the QCD corrected coefficient $ G$ at the $ u, d$ vertex on
the other hand.

The first one $ c_{1,2}$ are quantities that issue from the renormalization
group
equation that sums up large logarithmic enhancement due to gluons crossing
the $W$ line, i.e. gluons connecting the initial ${\bar s} c$ to the final
$ u {\bar d} $. This is why $ \tau $ decay is not concerned with this effective
Lagrangian ( gluons ignore $\tau $ lepton).}, then both $R_{D_s}$ and $ R_\tau$
in Eqs.(1) and (5) respectively have to be multiplied by a common correction
factor\footnote{
The QCD corrected coefficient $ G$ in Eq.(7), on the other hand, operates only
at the final state $u, d$ vertex, therefore $ G$ is common to both $R_\tau$ and
$R_{D_s}$.} $G$ given by \cite{7}
\begin{equation}
G~=~ 1+{\alpha_s \over \pi} \delta_1(x_u, x_d) + ({\alpha_s \over \pi})^2
\delta_2(x_u, x_d) + ({\alpha_s \over \pi})^3 \delta_3(x_u, x_d)
\end{equation}
with
\begin{equation}
 \delta_1(0,0) =1 ~,~ \delta_2(0,0)=5.2~,~ \delta_3(0,0)=26.36
\end{equation}
In Eq.(7) non-perturbative QCD contributions \cite{8} turn out to be tiny and
are consequently neglected.

For non-zero arguments $ x_u, x_d $,  the one-loop function $ \delta_1(x,y)$
has also been computed \cite{9} with a rather surprising feature $
\delta_1(x,y) > \delta_1(0,0)$  recently confirmed \cite{10}.
To our knowledge, $\delta_2(x, y)$ and $\delta_3(x, y)$ for non-zero $ x, y$
are not yet computed.

Now the first question that arises is which mass - current algebra mass or
constituent mass - must be used for the up, down quarks?

We argue for the first
one ( current mass) due both to theoretical and phenomenological reasons.
Theoretically, in QCD, the divergence of the observable axial weak current is
given in terms of current mass, as extensively discussed in Ref.[11]. The
constituent mass is only appropriate for the bound state problem not considered
here. Let us also remark that the hadronization of $u,~d$ quarks into pions is
conceptually different from the boundstate problem.

 At the phenomenological level, the use of constituent mass $\simeq 300~
MeV$ would firstly spoil the excellent arguments in favour of QCD tests
in $\tau$ decays \cite{7,8,12}:
indeed, the enhancement of about $ 20\%$ through the QCD factor $ G$
in Eq.(7) would be diminished by a similar amount if constituent mass is used
because of the phase space factor $ F(x,y)$ of Eq.(6). Secondly, for
constituent mass, the ratio $ R_{D_s}$ of Eq.(1): would be in strong
disagreement with data by two orders of magnitude.

Once the principle of current algebra masses is retained, let us return to
Eq.(1) and ask ourself which numerical value must be used for $ m_u$ and $
m_d$.

This is because, contrary to $ R_\tau$ in Eq.(5) which is insensitive to exact
values for current masses, the ratio $ R_{D_s}$ as given by Eq.(1) is very
sensitive to them.

Light quark masses have a reputation of being `` not well measurable"
\cite{11,13} and even their indirect experimental determinations have so far
not even been attempted \cite{11}, while there exists a huge number of
theoretical estimates \cite{13} ranging between $ 4~ MeV$ and $ 12~MeV$ for the
average mass $ {\bar m}$ of the $ u, d$ quarks.
\begin{equation}
{\bar m} = {1 \over 2} ( m_u + m_d)
\end{equation}
Since consensus holds \cite{13} for the ratio $ r = {m_d -m_u \over m_d+ m_u}
=0.28\pm 0.03$, the problem resides only on ${\bar m}$, more precisely on the
running mass ${\bar m}(s_0)$ at the scale $ s_0$ of few $GeV^2$.

Returning to the left-hand side of Eq.(1), we first estimate the experimental
value
of $R_{D_s}$ from the PDG \cite{PDG} to be $ 2.8\pm 1.2$ where the branching
ration $ Br(D^+_s \rightarrow X_{ud}) = (1.65\pm 0.35) 10^{-2}$ has been used
together with $ Br(D^+_s\rightarrow \mu^+ \nu)=(0.59\pm0.22) 10^{-2}$.

Within the one standard deviation limit of data for $R_{D_s}$, Eqs.(1), (7) can
only be satisfied with ${\bar m}$ around $38~MeV$ - a huge value much larger
than those commonly expected \cite{13}. In principle, there is nothing wrong
with
such a large ${\bar m}$ value, although this would imply a considerably lower
value for the $ -< {\bar \psi } \psi > $ condensate than the standard chiral
perturbation theory could support. Therefore before taking seriously this
crude $38~MeV$
value, we must ask ourself the next question, what could be the other sources
that may contribute to the substantially observed inclusive branching ratio
$Br(D_s\rightarrow X_{ud})$ ?

There exists at least two possibilities. The first one is via the so-called
Zweig forbidden rule as depicted in Fig.1: the dominant color-favoured decay
mode $ c\rightarrow s+(u{\bar d})$ when
combined with the spectator ${\bar s}$ could induce, through a two-gluon box
diagram, the chain $c{\bar s}\rightarrow ( s{\bar s}) +(u{\bar d}) \rightarrow
(q{\bar q}) + ( u{\bar d})$ where $ q$ stands for $ u$ and $ d$ quarks. It is
the final state interaction at the quark level (appropriated to inclusive
processes), in which the final state $ s{\bar s}$ turns into $ q{\bar q}$.

The second one, as depicted in Fig.2, is a weak annihilation accompanied by a
gluon emitted from the initial ${\bar s}$ and $ c$ quarks bound inside $ D_s$ ~:
mechanism, first proposed in Ref.\cite{14} to vitiate the helicity suppression
and recently being reexamined in details \cite{15}, could {\it \`a priori}
yield a large $D_s\rightarrow X_{ud}$ rate ( since the width is no longer
suppressed by $ {{\bar m}^2\over M^2}$).

Our task now is to compute these two contributions of Fig.1 and Fig.2, that we
called respectively final state strong transition (FT) and gluonic weak
annihilation (GA).

1) For the first one (FT), the decay rate can be written in the form

\begin{equation}\label{10}
\Gamma_{FT}(c{\bar s}\rightarrow (q{\bar q})+(u{\bar d}))=
\Gamma_{SP}(c\rightarrow
s+(u{\bar d}))P(s{\bar s}\rightarrow gg \rightarrow q{\bar q})
\end{equation}
where $\Gamma_{SP}(c\rightarrow s +(u{\bar d})) $ is the familiar spectator
inclusive
rate and $ P(s{\bar s} \rightarrow gg\rightarrow q {\bar q})$ is the transition
probability for the $s{\bar s}$ pair in the final state transforming
into a $ q{\bar q}$ pair through a box diagram.

The first term in Eq.(\ref{10})
$\Gamma_{SP}( c\rightarrow s u{\bar d})$ is given by:
\begin{equation} \label{11}
\Gamma_{SP}( c \rightarrow s+(u{\bar d})) = 3a^2_1{G^2m_c^5 \over 192 \pi^3}
|V_{cs} V_{ud}|^2 I(x_s, x_u, x_d)[1-{2\alpha_s \over 3\pi}(\pi^2-{31 \over
4})]
\end{equation}
In Eq.(\ref {11}), the phase space factor $I(x,y,z)$ is known \cite{6} with
$ I(x,0,0)=
1 -8x+8x^3-x^4-12x^2lnx $. The QCD correction factor $ C= -{2\alpha_s\over
3\pi}(\pi^2-{31\over 4})$ corresponds only to massless $s,~u,~d$ quarks and can
be decomposed into two parts, the ``upper vertex" associated to the $c{\bar s}$
part is $-{2\alpha_s\over 3\pi}(\pi^2-{25\over 4})$ and the ``lower vertex"
associated to $ u{\bar d}$ part is $ + {\alpha_s \over \pi}$.
For massive $ s,~u,~d $ quarks,
both "upper" and "lower" parts have already been computed \cite{9}, the
explicit analytic expression for the ``upper" part is also given \cite{10,16}.

We now compute the box diagrams ( there are two, with crossed gluons). The
dimensionless box diagram amplitude corresponding to
$ s(p_1)+{\bar s}(p_2)\rightarrow q(p_3)+{\bar q}(p_4)$ can be
conveniently written as:
\begin{equation}
{\cal A} =\Big ({4\alpha_s \over 3}\Big )^2 \hspace{2mm}
{{\bar v}(p_2)\gamma^\mu u(p_1)
{\bar u}(p_3)\gamma_\mu v(p_4) \over Q^2} B(Q^2, t, m_s^2)
\end{equation}
with
\begin{equation}\label{13}
Q^2=(p_1+p_2)^2 ~,~ t=(p_1-p_3)^2
\end{equation}
The dimensionless quantity $ B(Q^2, t, m_s^2)$ coming from loop integration has
the following representation ( we neglect $m_q^2$ )
\begin{equation}\label{14}
B(Q^2, t, m_s^2)=  \int^1_0dx\int^1_0dy\int^1_0dz
z(1-z){ Q^2[ Q^2 x(1-x)(1-z)^2-t(1-z)-m_s^2yz^2] \over
D^2( x, y, z, Q^2, t, m_s^2 ) }
\end{equation}
where
\begin{equation}\label{15}
D(x, y, z, Q^2, t, m_s^2) = - Q^2x(1-x)(1-z)^2 - ty(1-y)z^2 +m_s^2 y z^2
\end{equation}
Its explicit expression is given by:
\begin{eqnarray}\label{16}
B(Q^2, t, m^2_s)& = &ln^2(-{\eta_1 \over \eta_2})
+ {1\over \xi_1-\xi_2} {\cal
L}(Q^2,t,m_s^2) \nonumber \\
&+& {Q^2\over Q^2+t}\Big \{{Q^2+2t+2m_s^2 \over 2
(Q^2+t)}\big [Li_2(1+{t\over m_s^2}) -{\pi^2\over 6} - ln^2(-{\eta_1\over
\eta_2})\big ]
\nonumber \\
&+& {1\over2} Li_2(-{t\over m_s^2}) + {\eta_2-\eta_1\over 2} ln(-{\eta_1\over
\eta_2})\nonumber\\
& + & {Q^2+2t -4m^2_s{t\over Q^2} + 2 {m^4_s\over t} -2 {m_s^4 \over Q^2}
\over 2(Q^2+t)(\xi_2 - \xi_1)} {\cal L}(Q^2, t, m_s^2)\big \}
\end{eqnarray}
with
\begin{eqnarray}
{\cal L}(Q^2, t, m^2_s) &=& Li_2({\xi_1 \over \xi_1-\xi_2}) +
Li_2({\xi_1 \over \xi_1-\eta_2}) - Li_2({\xi_2 \over \xi_2-\eta_1})
- Li_2({\xi_2 \over \xi_2-\eta_2})\nonumber \\
2\eta_{1,2} &=& 1 \pm \sqrt{1-{4m_s^2 \over Q^2}} \nonumber \\
2\xi_{1,2} &=& 1 \pm \sqrt{1-{4m_s^2 \over Q^2}(1+{m^2_s \over t})} \nonumber
\end{eqnarray}
where $ Li_2$ is the Spence or dilogarithmic function.

Once the box amplitude ${\cal
 A}$ for $ s{\bar s} \rightarrow q{\bar q}$ is
known, the transition probability $ Y\equiv P(s{\bar s}\rightarrow gg
\rightarrow q{\bar
q})$ in Eq.(10) is then obtained by:
\begin{equation}\label{17}
Y = {1\over 4\pi^2} \int {d\vec p_3 \over 2 E_3} \int {d\vec p_4 \over 2 E_4}
\delta^4(Q-p_3-p_4)|{\cal A}|^2
\end{equation}
Since $  \int {d\vec p_3 \over 2 E_3} \int {d\vec p_4 \over 2 E_4}
\delta^4(Q-p_3-p_4) ={\pi \over 2}$, the quantity $ Y$ can be conveniently
rewritten as
\begin{equation}\label{18}
Y= \Big ({4\alpha_s \over 3}\Big )^4 {1\over \pi} {\overline {\cal A}
(Q^2, m^2_s)}
\end{equation}
The range of variation for $Q^2$:
\begin{equation}\label{19}
4m^2_s \leq  Q^2 \leq {m_s \over m_c}(m_s+m_c)^2
\end{equation}
is deduced from $ 0\leq l^2 \leq (m_c-m_s)^2$ where $ l$ is the invariant
mass of the $ u{\bar d}$ pair issuing from the spectator mechanism
$ c\rightarrow s+( u{\bar d})$.

The angular integration over the variable $ t $ in Eq.(\ref{17}) is done
numerically, and $ {\overline {\cal A}(Q^2, m_s^2)}$ turns out to be inside
$ 0.36 \leq {\overline {\cal A}}(Q^2, m_s^2) \leq 0.95 $ for $ Q^2$ inside
Eq.(\ref{19}), using $m_s=0.15~GeV,~ m_c=1.45~GeV$.

With these values put into Eqs.(\ref{10}), (\ref{11}) and (\ref{18}),  the FT
contribution yields a branching ratio $ Br(D_s \rightarrow~ pions~)_{FT} \leq
0.68\% $ which is still far from the experimental data, which at least equals
to
 $(1.65\pm0.35) \% $.

2) For the gluonic weak annihilation (GA) of Fig.2, the amplitude can be
directly taken from Eq.(4) of Ref.\cite{14} with the following substitution:
\begin{equation}
q\Rightarrow k_1+ k_2 ~~,~~ g_s\epsilon^\nu_j(q) \Rightarrow g_s^2 {\bar
u}(k_1) \gamma^\nu
{\lambda_j \over 2} v(k_2) {1\over (k_1+k_2)^2} ,
\end{equation}
where $ k_1$ and $ k_2$ are the $ q$ and ${\bar q}$ momentum, and $ j$ denotes
the color index.

We have
\begin{equation}
{\cal A} = {g^2_s G|V^*_{cs} V_{ud}|\over \sqrt{2}}{ R_{\mu\nu}( p,
k_1+k_2)\over (k_1+k_2)^2}~ {\bar u}(q_1)\gamma^\mu(1-\gamma_5)v(q_2) {\bar
u}(k_1) \gamma^\nu {\lambda \over 2} v(k_2),
\end{equation}
with
\begin{equation}
R_{\mu\nu}( p, q) ={M \over (p.q)}\big\{ F_A( q_\mu p_\nu - ( q.p)g_{\mu\nu})
+ i F_V \epsilon_{\mu\nu\alpha\beta} p^\alpha q^\beta \big \},
\end{equation}
where $ F_{V,A}$ are the two dimensionless form factors somehow reflected the
wave function at the origin of the $ c{\bar s}$ bound state, which in turn is
proportional to the decay constant $ f_{D_s}$ via the Van Royen- Weisskopf
formula. The tensor $ R_{\mu\nu}$ is reminiscent of the standard fermionic
triangular loop describing $ W^+ \rightarrow \gamma \pi^+ $ or $ \pi^o
\rightarrow \gamma \gamma $, taken as an example.

Here $ p= q_1 + q_2 + k_1 + k_2 $ with  $ p^2 = M^2 $. Neglecting $m_{u,d}^2$
compared to $ M^2$, the four-body phase space integration of ( spin and color
summing up) $|{\cal A}|^2 $ can be simplified, and the decay rate is computed
to be:
\begin{eqnarray}
\Gamma_{GA}( D_s \rightarrow u{\bar d} + q {\bar q})& = &a_1^2 {G^2 |V_{cs}^*
V_{ud}|^2 \alpha^2_s ( F_V^2+F_A^2) \over 12 \pi^3 M }
\int^{M^2}_0 dt_1 t_1 ( M^2-t_1 )^2     \nonumber \\
&& \int^{M^2}_{t_1} dt_2{(t_2-t_1)(M^2-t_2) \over t^2_2(M^2-t_1+t_2)^2},
\end{eqnarray}
with $ t_1=(q_1+q_2)^2 ~, ~ t_2 =(k_1+k_2)^2 $.

The double integration over $ t_1, t_2$ can be done analytically and the result
is
\begin{equation}
\Gamma_{GA}( D_s \rightarrow u{\bar d} + q {\bar q}) = a_1^2 {G^2 |V^*_{cs}
V_{ud} |^2 \alpha^2_s ( F_V^2 + F_A^2) M^5 \over 12 \pi^3} ( {\pi^2\over 6}
+{10 \over 3} ln2 - {71 \over 18}).
\end{equation}
Following Ref.\cite{14}, we will take
\begin{equation}
F_V={ f_{D_s} \over \sqrt{12} m_s} ~~, ~~ f_A = {f_{D_s}( m_s- m_c) \over
\sqrt{12} m_s m_c}.
\end{equation}
Numerically, it turns out that the $ \Gamma_{GA}$ rate is extremely small, the
corresponding branching ratio $ Br( D_s \rightarrow pions)$ via gluonic weak
annihilation mechanism is at most $(0.1)\%$. In Eqs.(12) and (24) we take
$\alpha_s=0.3$.

Putting altogether both $FT$ and $ GA$ contributions, the inclusive branching
ratio $ Br( D_s \rightarrow pions)$ due to these gluonic processes cannot
exceed $ 0.8\%$ and is still far from the observed inclusive branching ratio
larger than $(1.65\pm 0.35)\%$ as estimated from PDG \cite{PDG}.

The difference  between these two numbers must be attributed to the pure $ W$
annihilation tree diagram \footnote{ The $ W $ annihilation and the two gluonic
contributions have different decay products, therefore they contribute
incoherently to the $ D_s\rightarrow pions $ rate, without interference.}, from
which at the one
standard deviation lower bound of data, we get ${\bar m} \simeq 22~MeV$ using
again Eqs.(1) and (7).

This value, although smaller than the crude $ 38~MeV$ obtained
above ( when the gluonic backgrounds are neglected)
is still at least twice as large as
the common estimates.

\section{ Exclusive three pion mode: Mass determination from QCD sum rules}

We now consider the exclusive decay mode $ D^+_s \rightarrow \pi^+\pi^+\pi^-$
which allows us to  extract the spectral function $\rho(Q^2)$ that will be in
turns
exploited in the QCD sum rules to obtain ${\bar m}$.

The starting point is the QCD sum rule \cite{17} for the two-point correlator
of the divergence of the axial-current, put in the form \cite{11}:
\begin{equation}\label{20}
{\bar m}^2(s_0) = H^{-1}(w,s_0)\int^\infty_0dQ^2 w(Q^2, s_0) \rho(Q^2)
\end{equation}
We also consider its finite energy version closely followed Ref.\cite{18}
\begin{eqnarray}\label{21}
{\bar m}^2(\mu, s_0) & \equiv & \Big({ln s_0/\Lambda^2 \over
ln\mu^2/\Lambda^2}\Big )^{{24\over 29}} {4\pi^2 \over 3 s_0^2}
\big [ 1+ R_2(s_0) + 2c_4
<O_4>/s^2_0~~ \big ]^{-1}\nonumber\\
& &\{f_\pi^2 m_\pi^4 + \int^{s_0}_0 dQ^2 \rho(Q^2)\}
\end{eqnarray}
In Eq.(\ref {20}), $ w(Q^2,s_0)$ denotes weight function and $H(w,s_0)$ is
defined by the large $ Q^2$ behavior of the two-point correlator, it has a
perturbative QCD part and a non-perturbative part parametrized in terms of
vacuum condensates. Also the two-loop expression for $ R_2(s_0)$ as well as the
dimension-4 condensate $ c_4<O_4>$ in Eq.(\ref{21}) can be found in
Refs.\cite{18,22}.

These types of sum rules have been extensively used in the literature to
estimate quark mass and as explained in Ref.\cite{11}, the main problem in
these estimates is not the sum-rule technique itself but rather the
{\it complete absence of experimental information} on the magnitude of
$\rho (Q^2)$ beyond the one-pion contribution. The later $f_{\pi}^2 m_\pi^4 $
is singled out in Eq.(\ref{21}) with $ f_\pi \simeq 132~MeV$. Keeping only
this pion contribution, one finds \cite{11,18} a lower bound ${\bar m}(1~GeV)
> 4(5)~MeV$ due to the positivity of $ \rho(Q^2)$.

Fortunately, we show that the decay mode $ D_s \rightarrow 3\pi$ provides
precious
information on $\rho(Q^2)$ at fixed $ Q^2 = M^2$. Indeed, the decay amplitude $
D_s^+ \rightarrow \pi^+\pi^+\pi^-$ is given by
\begin{equation}
{G\over \sqrt{2}}V^*_{cs} V_{ud} a_1 f_{D_s} Q_\mu < \pi^+(p_1)
\pi^+(p_2)\pi^-(p_3)|A^\mu|0>
\end{equation}
where the most general matrix element $<3\pi| A^\mu| 0>$ can be expressed in
terms of three form factors \cite{19}, the same occurred in $ \tau \rightarrow
\nu + 3 \pi$. However, for $ D_s \rightarrow 3 \pi$, only one dimensionless
form factor \cite{20} associated to the divergence of the axial current is
involved and will be denoted by $ F(s_1, s_2)$
\footnote{ Our dimensionless form
factor $F(s_1, s_2)$ is related to the $F_4$ form factor
 of Ref.\cite{19} by $ F(s_1,s_2)= M F_4(s_1, s_2, Q^2=M^2)$}:
\begin{equation}\label{23}
Q_\mu<\pi^+(p_1)\pi^+(p_2)\pi^-(p_3) | A^\mu |0> = M F(s_1, s_2)
\end{equation}
with $ s_i= (Q-p_i)^2~,~ i=1,2$.

In terms of $ F(s_1, s_2)$, the $D_s \rightarrow 3\pi$ rate is given by
Ref.\cite{20}:
\begin{equation}\label{24}
\Gamma(D_s\rightarrow 3\pi) = a^2_1 {G^2 |V^*_{cs} V_{ud}|^2 \over 128 \pi^3}
M^3 f^2_{D_s} K  ,
\end{equation}
where the dimensionless constant $ K$ is obtained by integrating the squared
form factor $ | F(s_1, s_2)|^2 $ over the whole Dalitz domain:
\begin{equation}\label{25}
K \equiv {1\over 4 M^4}\int \int ds_1 ds_2 |F(s_1, s_2)|^2
\end{equation}
\begin{eqnarray}\label{26}
4m_\pi^2& \leq s_1 \leq & (M-m_\pi)^2 \nonumber\\
s_{min}(s_1) & \leq s_2 \leq & s_{max}(s_1)
\end{eqnarray}
$$
s_{min,max}(s_1) = {M^2+s_1-m^2_\pi \over 2} \pm \sqrt{{s_1-4m^2_\pi \over
s_1}} {\lambda(M^2, m_\pi^2, s_1) \over 2} .
$$
The numerical value of $ K$ that we can extract from Eq.(\ref{24}) has two
sources of errors, the first one is related to the experimental errors of the
non-resonant branching ratio $ Br(D^+_s \rightarrow \pi^+\pi^+\pi^-)_{NR} =
(1.01 \pm 0.35)
10^{-2}$, the second one is the uncertainty of the decay constant $f_{D_s}$.
Fixing $ f_{D_s} = 280~MeV$\cite{PDG}, we get \cite{20} $ K=0.486\pm 0.168$
where errors come from the ones of the experimental $D_s \rightarrow
\pi^+\pi^+\pi^-$ branching ratio. Let \cite{PDG} $ f_{D_s} = 280 \pm 70 ~MeV$,
then we have:
\begin{equation}\label{33}
0.27\pm 0.06 \leq K \leq 0.74 \pm 0.17
\end{equation}
Now the crucial point is that the constant $ K$ can be directly related to the
spectral function $ \rho_{3\pi}(Q^2)$ entering in the QCD sum rule
Eqs.(26),(27). At $ Q^2=M^2 $, we get:
\begin{equation}\label{34}
\rho_{3\pi}(Q^2=M^2) = {K\over 64 \pi^4} M^4 .
\end{equation}
Eq.(34) is obtained if we remind that the spectral function is defined by
\begin{equation}\label{35}
\rho(Q^2)={1\over 2\pi}\sum_n (2\pi)^4
\delta^4(Q-p_n) |<n|\partial_\mu A^\mu |0>|^2
\end{equation}
and in particular for $\rho_{3\pi}(Q^2)$:
\begin{equation}\label{36}
\rho_{3\pi}={1 \over 2 \pi}\int\int\int {d\vec p_1 \over 2E_1}
{d\vec p_2 \over 2E_2}      {d\vec p_3 \over 2E_3} {1\over (2\pi)^9} (2\pi)^4
\delta^4(Q-p_1-p_2-p_3)|<3\pi|A^\mu|0>Q_\mu|^2
\end{equation}
In Eqs.(33) and (34), the constant $K$ and consequently the
$\rho_{3\pi}(M^2)$ are only related to the charged $ D^+_s\rightarrow
\pi^+\pi^+\pi^-$ mode. The neutral mode $ D^+_s \rightarrow \pi^0\pi^0\pi^+$ by
isospin consideration \cite{20} is presumably   $1/4$ of the charged one.
Hence the total charged and neutral must be $5/4$ of Eq.(\ref{34}).

As shown in Eqs.(30) and (34), from $ D_s\rightarrow 3\pi$ data, we get
information for the spectral function $\rho(Q^2)$ at only one fixed value of $
Q^2=M^2$, while the QCD sum rules Eqs.(26) or (27) need the whole
range of $ Q^2$ in $ \rho(Q^2)$. Fortunately, we also know that at large $Q^2$,
perturbative QCD gives \cite{18} explicit analytic expression for $\rho(Q^2)$
in terms of one unknown parameter $(\hat m )^2$:
\begin{equation}\label{37}
\rho(Q^2) \rightarrow {3 \over 2\pi^2} {(\hat m)^2 Q^2 \over ({1\over 2}
lnQ^2/\Lambda^2 )^{{12\over 29}}}[ 1+{17\over 3}{\alpha_s(Q^2) \over \pi}]
\end{equation}
Since the normalization at $ Q^2=M^2$ for $\rho_{3\pi}(Q^2=M^2)$ is fixed by
${5 K\over 256\pi^4} M^4 $, let us follow the usual procedure \cite{21}
by adopting the simplest duality ansatz parametrization for $\rho(Q^2)$:
\begin{equation}\label{38}
\rho(Q^2)={5 K\over 256 \pi^4} M^2 Q^2 ({lnM^2/\Lambda^2 \over ln
Q^2/\Lambda^2})^{{12\over29}}{1+{17\over 3}{\alpha_s(Q^2)\over \pi} \over
1+{17\over 3}{\alpha_s(M^2)\over \pi}}
\end{equation}
which reduces to $ \rho(M^2)={5 KM^4 \over 256 \pi^4}$ as it should be.

Putting now Eq.(38) into the sum rule Eq.(26), and taking for $K$
the one standard deviation most conservative lower limit in Eq.(33), we
then obtain for ${\bar m}(1~GeV)$ a quite large value $ 20~MeV$, consistent
with our previous result from the inclusive decay $ D_s \rightarrow X_{ud}$.

A different parametrization for $\rho(Q^2)$ can be obtained by making use of
the shape of the low energy end of the $3\pi$-threshold $\rho^{3\pi}_{\chi}(t)$
 calculated in a recent paper \cite{22} using lowest order chiral
perturbation theory. Following Ref.\cite{22}, we will make use of the finite
energy sum rule Eq.(27) with $\rho(Q^2)$ given by
\begin{equation}\label{39}
\rho(Q^2) = \lambda{m^4_\pi \over 64\pi^4 f^2_\pi}{Q^2 \over 18}\rho_{had}(Q^2)
\end{equation}
where $ \rho_{had}(Q^2)$ is taken from Eq.(32) of Ref.\cite{22} encoding
the presence of two $ \pi'$ resonances, the $ 1300 ~MeV$ and $ 1770~MeV$.

Here $\lambda$ is an overall normalization which, as emphasized by the authors
of Ref.\cite{11}, can only be determined by experimental informations and not
by theoretical calculations. Matching Eq.(34) to Eq.(39) at
$ Q^2 =M^2 $, one can fix $\lambda$. It turns out that our overall
normalization $\lambda$ provided by $ D_s\rightarrow 3\pi $ decay rate is
much larger than the one given in Ref.\cite{22}, and that
is the crucial reason for getting high value of $ {\bar m} $.

\section{Summary and Conclusion}
The starting point is our observation that the decay rate of the
charmed $ D_s$ meson into pions is surprisingly larger than common
estimates. The amplitude
being governed by the $ W$ annihilation is usually expected to be negligible
by the partial conservation of the axial current ( PCAC) at relatively high
energy.

We then go beyond the tree $W$ annihilation diagram and consider two possible
contributions. The first one, called finite state strong transition ( FT)
coming from the dominant spectator decay
$c {\bar s} \rightarrow s{\bar s} + u {\bar d}$ followed by the Zweig violating
rule $ s{\bar s} \rightarrow q{\bar q}~~ ( q= u, d)$ via the two-gluon box
diagram. The second one, called gluonic weak annihilation ( GA), is a genuine
weak annihilation accompanied by an emission of one gluon from the quarks
$ c, {\bar s} $ bound inside $ D_s$, the mechanism that invalidates the
helicity suppression.

 We find out that these gluonic effects together cannot explain the large
inclusive $ D_s$ decay into pions, and consequently we suggest that $ W$
annihilation (tree diagram) also
contributes substantially to the rates. This would imply that ${\bar m}$, the
averaged $ u,~d$ mass, could be around $ 22~MeV$, few times as large as the
usual estimates \cite{13}.

One might argue that our quark parton type of analysis for inclusive decays of
charm could only be accurate within a factor of two. However, by analyzing the
exclusive $ D_s\rightarrow 3\pi$ mode, we obtain
the normalized spectral function $ \rho_{3\pi}(Q^2)$, which is in turns
exploited in the QCD sum rules.
Again, by this completely different method, a higher value  than
common estimates is obtained for  ${\bar m}$.

Our lower bound of $ 22~MeV $ for ${\bar m}$ depends, on the one hand, on the
experimental data of both inclusive and exclusive $D_s$ decays into pions, and
on the other hand, on the theoretical factorization method \`a la BSW.
To obtain such a mass value, we carefully take into account
experimental errors as well as the uncertainty of the decay constant $
f_{D_s}$.

Independent of our proposition for a high ${\bar m}$ as a solution to the
substantial rates of $D_s$ into pions, the understanding of the origin of such
data constitutes a very interesting problem in its own right.
The confirmation of experimental data is equally important,
on the other side.

Finally we would like to emphasize the similarity as well as the
complementarity between $\tau $ lepton and charmed $D_s$ meson decaying into (
three and more) pions: While $\tau $ probes the dominant spin $1$ and the small
spin $0$ axial spectral functions in the whole range of the momentum transfer $
0\leq Q^2 \leq M^2_\tau $ ( to separate them is a challenging
experimental problem
\cite{11}), the $ D_s$ probes directly and easily the spin $0$ spectral
function at only one value $ Q^2=M^2_{D_s}$. This latter is the key for the
determination of the light quark mass, a fundamental parameter `` not well
measured" in the standard model. Its importance in chiral symmetry breaking
from both perturbative and nonperturbative aspects is well-known.

\newpage
\vspace{5mm}
\hspace{1cm} \large{} {\bf Acknowledgements}    \vspace{0.3cm}
\normalsize

\vspace{2mm}
One of us (P. X. Y.) would like to thank  Y. Y. Keum, J. Stern and
T. N. Truong for helpful discussions.


\newpage
\begin{figure}
\end{figure}

\begin{figure}
\end{figure}

\end{document}